\documentclass[review,1p]{elsarticle}

\usepackage{lineno,hyperref}
\modulolinenumbers[5]

\newcommand{\ToolNameNoSpace}{\texttt{Griffin}}
\newcommand{\ToolName}{\ToolNameNoSpace~}

\usepackage{listings, xcolor}

\definecolor{verylightgray}{rgb}{.97,.97,.97}

\lstdefinelanguage{Solidity}{
	keywords=[1]{anonymous, assembly, assert, balance, break, call, callcode, case, catch, class, constant, continue, constructor, contract, debugger, default, delegatecall, delete, do, else, emit, event, experimental, export, external, false, finally, for, function, gas, if, implements, import, in, indexed, instanceof, interface, internal, is, length, library, log0, log1, log2, log3, log4, memory, modifier, new, payable, pragma, private, protected, public, pure, push, require, return, returns, revert, selfdestruct, send, solidity, storage, struct, suicide, super, switch, then, this, throw, transfer, true, try, typeof, using, tx.value, view, while, with, addmod, ecrecover, keccak256, mulmod, ripemd160, sha256, sha3}, % generic keywords including crypto operations
	keywordstyle=[1]\lst@ifdisplaystyle\color{blue}\bfseries \fi,
	keywords=[2]{address, bool, byte, bytes, bytes1, bytes2, bytes3, bytes4, bytes5, bytes6, bytes7, bytes8, bytes9, bytes10, bytes11, bytes12, bytes13, bytes14, bytes15, bytes16, bytes17, bytes18, bytes19, bytes20, bytes21, bytes22, bytes23, bytes24, bytes25, bytes26, bytes27, bytes28, bytes29, bytes30, bytes31, bytes32, enum, int, int8, int16, int24, int32, int40, int48, int56, int64, int72, int80, int88, int96, int104, int112, int120, int128, int136, int144, int152, int160, int168, int176, int184, int192, int200, int208, int216, int224, int232, int240, int248, int256, mapping, string, uint, uint8, uint16, uint24, uint32, uint40, uint48, uint56, uint64, uint72, uint80, uint88, uint96, uint104, uint112, uint120, uint128, uint136, uint144, uint152, uint160, uint168, uint176, uint184, uint192, uint200, uint208, uint216, uint224, uint232, uint240, uint248, uint256, var, void, ether, finney, szabo, wei, days, hours, minutes, seconds, weeks, years},	% types; money and time units
	keywordstyle=[2]\lst@ifdisplaystyle\color{teal}\bfseries \fi,
	keywords=[3]{block, blockhash, coinbase, difficulty, gaslimit, number, timestamp, msg, data, gas, sender, sig, tx.value, now, tx, gasprice, origin, IRs, Expression},	% environment variables
	keywordstyle=[3]\lst@ifdisplaystyle\color{violet}\bfseries \fi,
	identifierstyle=\color{black},
	sensitive=false,
	comment=[l]{//},
	morecomment=[s]{/*}{*/},
	commentstyle=\color{gray}\ttfamily,
	stringstyle=\color{red}\ttfamily,
	morestring=[b]',
	morestring=[b]"
}

\usepackage{amsthm}
\newtheorem{definition}{Definition}
\usepackage{amsmath}
\usepackage{algorithm}
\usepackage{algpseudocode}
\usepackage[graphicx]{realboxes}

\usepackage{url}

\newcommand{\specialcell}[2][c]{%
	\begin{tabular}[#1]{@{}c@{}}#2\end{tabular}}

\journal{Journal of Systems and Software}

\begin{document}
\begin{frontmatter}
\title{Effective Targeted Testing of Smart Contracts}

\author[1]{Mahdi Fooladgar}
\author[2]{Fathiyeh Faghih}
\address{College of Engineering, University of Tehran, Tehran, Iran}
\address[1]{email: m.fooladgar@ut.ac.ir}
\address[2]{email: f.faghih@ut.ac.ir}

\begin{abstract}
Smart contracts are autonomous and immutable pieces of code that are deployed on blockchain networks and run by miners. They were first introduced by Ethereum in 2014 and have since been used for various applications such as security tokens, voting, gambling, non-fungible tokens, self-sovereign identities, stock taking, decentralized finances, decentralized exchanges, and atomic swaps. Since smart contracts are immutable, their bugs cannot be fixed, which may lead to significant monetary losses. While many researchers have focused on testing smart contracts, our recent work has highlighted a gap between test adequacy and test data generation despite numerous efforts in both fields. 
Our framework, \ToolNameNoSpace, tackles this deficiency by employing a targeted symbolic execution technique for generating test data. This tool can be used in diverse applications, such as killing  the survived mutants in mutation testing,
 validating static analysis alarms, creating counter-examples for safety conditions, and reaching manually selected lines of code.
This paper discusses how smart contracts differ from legacy software in targeted symbolic execution and how these differences can affect the tool structure, leading us to propose an enhanced version of the control-flow graph for Solidity smart contracts called CFG+. We also discuss how \ToolName can utilize custom heuristics to explore the program space and find the test data that reaches a target line while considering a safety condition in a reasonable execution time.
We conducted experiments involving an extensive set of smart contracts, target lines, and safety conditions based on real-world faults and test suites from related tools. The results of our evaluation demonstrate that \ToolName can effectively identify the required test data within a reasonable timeframe.
\end{abstract}

\begin{keyword}
smart contracts testing \sep symbolic execution \sep test data generation \sep Solidity
\end{keyword}
\end{frontmatter}

%\linenumbers

\section{Introduction}

In 2014, Ethereum introduced Solidity, a JS-like Turing-complete programming language for implementing and deploying smart contracts. Smart contracts are immutable autonomous pieces of code that are stored and deployed on Blockchain systems. Due to the distributed nature of blockchain, some of the Ethereum nodes, called miners, are responsible for executing these programs and storing the results. Smart contracts have been utilized for various applications, such as security tokens, voting, gambling, non-fungible tokens, lottery, property ownership, stock taking, decentralized finances, self-sovereign identity, decentralized exchanges, and atomic swaps.

Millions of dollars are currently being held by smart contracts, and any security or logical mistake in these contracts could result in significant financial losses. For example, the first bug in Ethereum resulted in the DAO attack~\cite{theDAO}, which caused a loss of \$150 million and led to the creation of a fork to reverse the attack. Since then, many other bugs have been discovered. The MonoX decentralized exchange recently suffered a \$31 million attack due to a logical error~\cite{Goodin2021}. This error caused a token's price to increase when it was swapped with itself.

Due to the immutable nature of blockchain, fixing the smart contracts bugs is not feasible. Therefore, it is crucial for smart contract owners and developers to test and verify their contracts thoroughly before they are deployed. 
Hence, various methods have been recently explored for testing and analyzing smart contracts in academic and technical works. Some focus on static analysis~\cite{mavridou2018tool,mavridou2018designing,josselinfeist2018, feist2019slither,tikhomirov2018smartcheck,remix-ide-static-analysis} that can identify known bugs and vulnerabilities in a smart contract by finding known bug patterns on the software, its control-flow-graph, or its abstract syntax tree. Despite its high performance and efficiency, this technique may generate numerous false positives and miss new bug patterns. Also, the static analysis tools can not generate test data that reveals the bugs. Another method is dynamic analysis (testing)~\cite{truffle-website,waffle-website,ellul2018runtime}. This method does not have the limitations mentioned for static analysis; however, it is slower and has it's own limitations on generating complete test suite. Different scenarios are designed to test a smart contract, and the smart contract is run under each specified scenario, typically in a simulated Ethereum virtual machine. The result of each scenario will be evaluated based on the expected result in a procedure called test oracle. If they are not matched, a failure is detected, and the root cause needs to be found. The most critical challenges in testing smart contracts are the following:
\begin{itemize}
	\item \textbf{Designing test oracles:} In the case of large number of test cases, it is not practical to manually evaluate the test results. On the other hand, it is not always straightforward to calculate the expected result of a test scenario to have an automated test oracle.
	\item \textbf{Measuring test adequacy:} One of the most important questions in the literature of software testing is how to measure the adequacy of test cases for evaluation of a software.
	\item \textbf{Automated generation of effective test data:} Assuming that a good measure for test adequacy is accessible, the next challenge is to automatically generate test cases, so that the test adequacy with respect to the measure is high enough.
\end{itemize}
Considering the second challenge, different criteria for measuring the test adequacy of smart contracts have been proposed. \textit{Coverage criteria} are the most known way of measuring test adequacy and include computing the number of statements, branches, and values the test suite covers. There are also more complicated coverage criteria; for example, MC/DC coverage checks if the test suite evaluates each clause in a complex condition as both true and false.
 Mutation testing is considered one of the most effective methods for evaluating the completeness of test design. This approach involves creating small syntactic changes, known as ``mutants," to the original program and testing them with a test suite. It is considered complete if the test suite has a different output in the original program and all its mutants. This technique is based on the hypothesis that experienced developers generally make minor syntactic errors. Thus, it aims to generate all possible faulty programs and evaluate whether the test suite is comprehensive enough to identify these errors.
 
Having criteria to measure test adequacy, the next challenge, as mentioned above, is to generate effective test cases accordingly. One popular approach is fuzzing~\cite{Wustholz}, which involves injecting invalid, malformed, or unexpected inputs into a system to reveal software defects and vulnerabilities. However, while fuzzing techniques have acceptable performance, they cannot achieve high coverage values. Another approach is symbolic execution~\cite{luu2016making,mythril-github, mossberg2019manticore,So2021}, which assumes inputs as symbols and tries to find the inputs that can reach each path in the program. These tools utilize SMT solvers to find and concretize the input values in an exhaustive search, while most aim for high branch coverage. Therefore, optimizing the balance between performance and achieving high coverage is crucial for these tools. It is worth mentioning that these tools cannot terminate for most real-size software.
 
In our recent study~\cite{fooladgar2021testsmart}, we aimed to evaluate the effectiveness of test suites generated by symbolic execution tools. To measure the completeness of the generated test data, we used mutation testing with a comprehensive set of regex-based mutation operators. Our results showed that symbolic execution tools such as Manticore~\cite{mossberg2019manticore}  have limitations in creating effective test suites, which can result in low mutation scores. Also, according to another recent study~\cite{chaliasos2024smart}, off-the-shelf tools can only prevent 8\% of attacks, which amounts to \$149 million out of the \$2.3 billion in losses. The study concluded that most logic-related bugs and protocol-related vulnerabilities are significant threats that are not adequately addressed by existing tools.
This findings highlights a gap between test adequacy and test generation fields despite numerous works in both areas. In other words, no tool is currently available to generate test data that matches the reachability condition for killing mutants in mutation testing for smart contracts. To bridge this gap, we propose using test adequacy metrics as a heuristic for test data generation.

We believe this solution can also connect static analysis with dynamic testing.
Despite the numerous publications on static analysis, some issues still need to be solved in the field. While static analysis tools aim to reduce false positives, developers still require a dynamic testing tool to confirm the accuracy of alarms, such as possible vulnerabilities, bad smells, logical errors, and programming issues generated by static analysis.

We have introduced the first targeted backward symbolic execution tool for Solidity smart contracts, called \ToolNameNoSpace, to address the above two issues. Our open-source tool can help software developers and testers generate test data for reaching any arbitrary line in the smart contract's source code. \ToolName is designed for various applications, including but not limited to:
\begin{itemize}
\item Generating test data that can reach a manually selected line of code 
\item Validating static analysis alarms and identifying input values that can cause software faults 
\item Determining the reachability condition for killing mutants 
\item Creating counter-example test data for any safety condition in a specific smart contract's source code line. 	
\end{itemize}

To achieve these objectives, \ToolName can utilize any heuristic to examine all possible paths from the entry points to the target line on our proposed enhanced version of Control Flow Graph, i.e., CFG+. In the prototype version, we introduced two heuristics: \textbf{(1)} a Floyd-Warshall-based heuristic that prioritizes exploring the shortest paths using the known Floyd-Warshall algorithm for finding shortest paths in graphs and \textbf{(2)} our proposed heuristic that analyzes the define-usage state of the state-variables to eliminate some infeasible paths early. We highlight the effectiveness of selecting an appropriate heuristic based on the smart contract and the target, as this can decrease the execution time of \ToolName.

The remainder of this paper is structured as follows. We will start by discussing testing methods and previous research on testing smart contracts in Section~\ref{sec:background}. Next, we will present our problem statement in Section~\ref{sec:problem}. After that, we will outline the main structure, design, and prototype implementation of our tool, \ToolNameNoSpace, in Section~\ref{sec:our-work}. The evaluation results will be presented in Section~\ref{sec:evaluation}. In Section~\ref{sec:related-works}, we will compare our tool with other related works and research. Finally, we will discuss the findings and future work on \ToolName in Section~\ref{sec:conclusion}.
\section{Background}
\label{sec:background}

In this section, we provide a brief introduction to the concepts used throughout the paper. We first give a brief introduction to Solidity. Then we talk about the bugs and vulnerabilities in smart contract. Following that, we give a brief introduction to symbolic execution and mutation testing as two techniques for increasing the dependability of smart contracts.

\subsection{Smart Contract}

Smart contracts are immutable pieces of code that are deployed and executed on blockchains. Once deployed, the code becomes immutable and is stored on the blockchain network. Any user on the network, including other smart contracts, can call its functions by initiating a transaction. Then, specific nodes on the network called ``miners" run the smart contract based on the transaction and post the verifiable result to the blockchain network. Due to the consensus protocol, it should be noted that all the network miners can verify the correctness of the posted results; hence, malicious miners cannot post incorrect results. Additionally, smart contracts are autonomous and perform based on the distributed nature of blockchain networks. The benefits of smart contracts have led to their increasing use in various applications, such as security tokens, voting, gambling, non-fungible tokens, lottery, property ownership, stock taking, decentralized finances, self-sovereign identity, decentralized exchanges, and atomic swaps.

Ethereum was the first blockchain to introduce a Turing-complete language called Solidity for developing smart contracts. Solidity, a programming language similar to JavaScript, allows Ethereum smart contract developers to write verifiable and autonomous smart contracts. The Ethereum smart contracts are stored in an Ethereum account, which holds the contract's compiled code (bytecode), specified data storage, and its balance in \textit{Ethers} (like regular accounts). In order to run smart contracts, Ethereum nodes must pay a fee, called \textit{gas}, to incentivize miners to verify and run the transactions. Gas payment also prevents wasteful tasks and infinite loops due to Turing-completeness.

Ethereum miners execute bytecodes on the Ethereum Virtual Machine (EVM). EVM is a stack-based virtual state machine with volatile memory and non-volatile account storage. The current state of each account is saved on non-volatile storage, which can store the smart contract's read-only source code, values for global variables, and the contract balance. Ethereum transactions can call any of a smart contract's declared functions, allowing the EVM to transition from one state to another by executing the related bytecode and changing the corresponding data storage section.

The bytecode of a smart contract is immutable and cannot be modified. It is loaded into a virtual read-only memory (ROM) during the execution of a transaction. A transaction must be executed to call the smart contract's constructor to deploy a smart contract on the Ethereum blockchain. This action results in storing the bytecode in the storage. It should be noted that the constructor can only be executed once.

The Ethereum Virtual Machine (EVM) is responsible for managing various global variables that smart contracts can use. The \lstinline|tx| variable stores details about the currently-executing transaction, such as the account that initiated it and the gas price. The \lstinline|msg| variable stores information about the previous call's data, including the caller, the name of the function that was called, the parameters passed to the function, and the amount of Ether sent to the smart contract. Finally, the \lstinline|block| variable contains information about the block being mined, such as the timestamp, the current gas limit for the network, the network difficulty, and the previous block. These global variables can be used by smart contracts in their source code. It is worth noting that, a smart contract can alter its storage, transfer Ethereum, call other contracts, and emit events during a transaction.

\subsection{Bugs and Vulnerabilities of Smart Contract}\label{sec:bugs-and-vuln-of-smart-contracts}
Nikolic et al. discovered that in 2018, there were over 34000 vulnerable smart contracts on the Ethereum network containing more than 4900 Ethers~\cite{nikolic2018finding}. Additionally, Certik\footnote{https://www.certik.com/} has reported over 60,000 vulnerabilities in smart contracts as of September 2023. Moreover, Soud et al. identified 2143 vulnerabilities from public coding platforms and sorted these vulnerabilities into 11 categories~\cite{soud2024fly}. Attackers have previously exploited these vulnerabilities, causing significant financial losses. The infamous DAO attack, which resulted in a loss of 150 million dollars, was the first of many subsequent attacks.  Zhang et al. have categorized the bugs into 17 categories: assertion failure, arbitrary writes, block-state dependency, compiler errors, control-flow hijack, Ether leak, freezing Ether, gas-related issues, integer bug, mishandled exception, precision loss, re-entrancy, suicidal contract, transaction-ordering dependency, transaction origin use, uninitialized variable, and weak PRNG~\cite{zhang2023demystifying}. Among them all, we provide more details on the most famous examples of these attacks.

In 2017, a vulnerability named the ``Parity Multi-Sig~\cite{parity-multisig-bug}" bug was discovered, which allowed anyone to take control of any multi-sig Parity wallet by calling the \lstinline|init| function on the parent contract using the fallback function on the child contract. Later that year, another similar vulnerability was found on Parity, known as the ``Accidental Kill~\cite{parity-accidentally-kill}" bug, where someone inadvertently called the \lstinline|initWallet| of the parent wallet directly, resulting in taking ownership of it. Then, he ran the  \lstinline|selfdestuct| and removed the parent contract from the Ethereum network. As a result, all Parity wallets on the Ethereum network were left orphaned, resulting in the loss of 300 million dollars.

Many other bugs have been revealed over the years on Solidity smart contracts. A token named RUNE in THORChain, a decentralized exchange, used \lstinline|tx.origin| instead of \lstinline|msg.sender|, which caused an 8 million dollar attack in July 2021~\cite{thorchain}. The bug was caused by the logic of \lstinline|tx.origin| statement in Solidity that if a chain of transactions has been received to a smart contract (e.g., $A \rightarrow B \rightarrow C$), the \lstinline|tx.origin| returns $A$ and \lstinline|msg.sender| returns $C$. Therefore, an attacker can convince you to call his contract, which calls the target contract, and lead \lstinline|tx.origin| to return the victim's address. Moreover, many software, including smart contracts, suffer from repetitive overflow and underflow bugs. Examples include the PoWH bug causing a loss of 800 thousand dollars~\cite{powh-underflow}, the BeautyChain bug affecting 100 million tokens~\cite{bec-batch-overflow}, and an ICON bug leading to a potential 800 million dollar loss~\cite{icon-fatal}.

Logical bugs are also crucial in Solidity smart contracts. In 2021, MonoX Finance decentralized exchange was the target of an attack that resulted in a loss of 31 million dollars~\cite{Goodin2021}. The attack was possible due to a logical bug in the MonoX contract that allowed input and output tokens to be identical in an atomic swap. As a result, when the attacker swapped the same tokens as an input and output, the increase in the output token's price overwrote the decrease in the input token. This caused the token's price to increase, leading to a significant loss.

Considering the above history of bugs and vulnerabilities, numerous studies have been conducted in the realm of smart contract analysis, which can be broadly classified into three primary categories; formal verification, static analysis, and dynamic testing. Formal verification is a method to check software mathematically. The software is described as a formal model, and the method examines if it satisfies a specific formal property. To achieve this, abstraction techniques convert the software into formal models. Later, model-checking approaches are applied to check the given formal property. However, for real-world software, the performance of these methods is typically low. Static analysis is a method used to analyze software in order to detect bugs and vulnerabilities without actually running the program. This is achieved by examining the program's source code, CFG, AST, data flow graph, and other graphs to check for known bug patterns. However, it's important to note that static analysis tools may generate false positive alarms, and they are not capable of identifying zero-day bugs and vulnerabilities. Therefore, dynamic testing techniques to analyze smart contracts are used as a complementary method. 

\subsection{Mutation Testing of Smart Contract}
There are two main categories of works in the field of dynamic testing; measuring the effectiveness of test suites, and test suite generation. 
Considering the challenge in measuring the effectiveness of test suites, different criteria for measuring the test adequacy of smart contracts have been proposed. \textit{Coverage criteria} are the most known way of measuring test adequacy and include computing the number of statements, branches, and values the test suite covers. There are also more complicated coverage criteria; for example, MC/DC coverage checks if the test suite evaluates each clause in a complex condition as both true and false.
Furthermore, mutation testing is considered one of the most effective methods for evaluating the completeness of test design. It involves making small syntactical changes to the source code, referred to as mutation operators. The concept behind this technique is that faults in software written by experienced developers are often due to tiny errors in syntax. For instance, a developer might accidentally use \lstinline|>| instead of \lstinline|>=|. To implement mutation testing, mutation operators are applied to the source code of the target program one at a time to create mutants. These mutants are then evaluated along with the original program by running a set of test cases (test suite). If the output of the program and a mutant differ on a test case, the mutant is killed, and the test case is selected. The chosen test cases can identify the difference between the program and its mutants, making them effective in identifying bugs.

In the literature, there have been numerous proposals for mutation testing in smart contracts.  SuMo~\cite{barboni2021sumo} is a tool developed by Barboni et al. that implements 44 mutation operators based on the latest Solidity documentation. The authors evaluated the effectiveness of these operators using two real-world smart contracts and associated test suites. The experimental results indicate that the mutation score for both contracts is moderately low. In another study, Chapman et al. suggested 61 mutation operators and tested them on three contracts and their corresponding test suites~\cite{chapman2019deviant}. Despite having high branch or statement coverage, the authors found that even well-designed test suites may result in low mutation scores. Also, some other similar works are trying to propose mutation operators such as \cite{honig2019practical,li2019musc} or even scalability of mutation testing in smart contracts~\cite{hartel2020mutation}.
Andesta et al.~\cite{andesta2020testing} conducted an extensive study on known bugs in smart contracts and proposed 49 Solidity-specific mutation operators across 10 classes based on these faults. Their study found that these operators could replicate 10 out of 15 of the most expensive real-world Solidity bugs.

\subsection{Symbolic Execution of Smart Contracts}

Several academic studies have been conducted to find ways to generate test suites for testing smart contracts. These studies fall into three main categories: symbolic execution, fuzzing, and search-based automatic test suite generation. Among these categories, symbolic execution tools are considered more powerful. Symbolic execution involves executing programs in an abstraction by considering inputs as symbolic and performing computations symbolically. Hence, symbolic execution tools can generate the necessary conditions for each path in the program and use SMT solvers to check if these paths can be reached using the inputs.

Oyente~\cite{luu2016making} is a symbolic execution tool that can be run on EVM byte-code version of smart contracts and tries to find known bug patterns such as timestamp dependency, re-entrancy, etc. Luu et al. evaluated their tool on more than 19000 Ethereum smart contracts in this work and found about 9000 known bugs and vulnerability patterns.
Mythril~\cite{mythril-github} is another symbolic execution tool for Solidity smart contracts, capable of detecting various vulnerabilities and bugs such as safety failures, overflow and underflows, unauthorized access to EVM storage, and bad smells on delegated calls. Harvey~\cite{Wustholz}, another tool from the Mythril developers, is a graybox fuzzer for Solidity smart contracts. In contrast to similar tools, this tool is able to generate multi-transaction test data to reach bugs and vulnerabilities.
Manticore~\cite{mossberg2019manticore} is a dynamic symbolic execution tool that is open-source and supports various formats, including EVM, Linux ELF binaries, and WASM modules. It emulates the Ethereum virtual machine symbolically, which enables it to represent transactions, storage, global values, and EVM's current state. Developers can interact with Manticore and even integrate it into their toolchain using the APIs provided.
SmarTest is also a symbolic execution tool that uses machine learning to learn the language model and a language-model-guided symbolic execution approach to generate a test suite of vulnerable transaction sequences~\cite{So2021}.

\section{Problem: Effective Test-data Generation}
\label{sec:problem}

\begin{figure*}
	\centering
	\includegraphics[width=\textwidth]{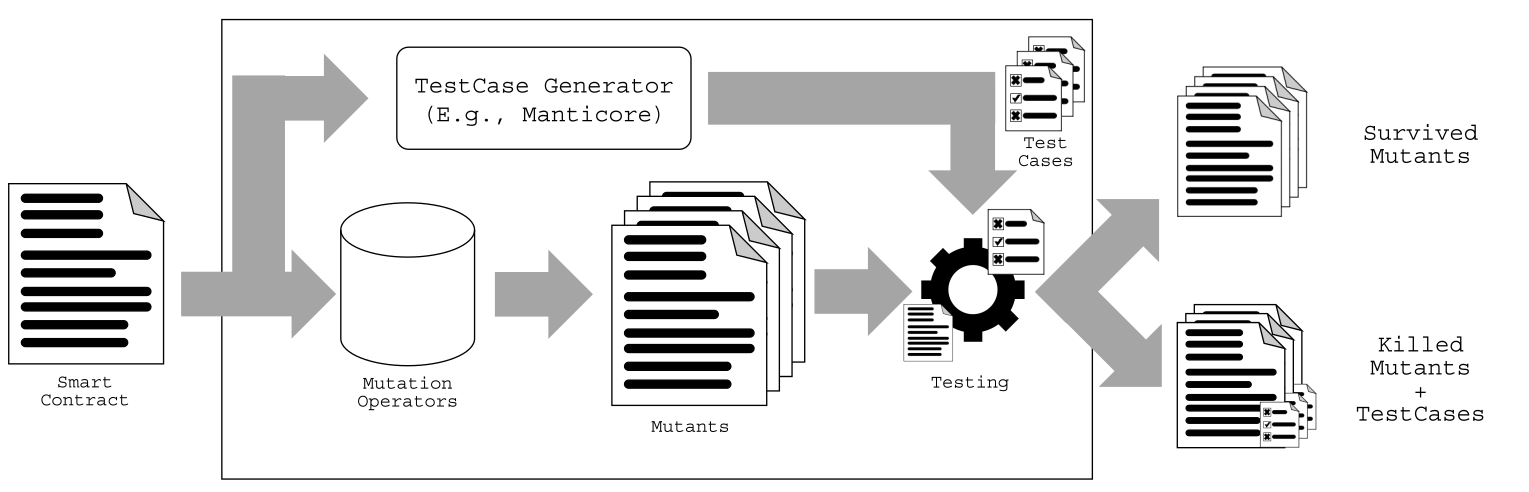}
	\caption{TestSmart Tool Structure~\cite{fooladgar2021testsmart}}\label{fig:testsmart-tool-structure}
\end{figure*}

%\subsection{The Survived Mutants Problem}
	Our previous work aimed to bridge the gap between generating test suites and selecting test cases through mutation testing. To that end, we developed a tool called TestSmart~\cite{fooladgar2021testsmart} as illustrated in Figure~\ref{fig:testsmart-tool-structure}. This tool inputs a Solidity smart contract and applies mutation operators to generate mutants. It then uses an automatic test-case generator tool like Manticore to generate the test data. Dynamic testing is subsequently performed on the mutants to produce a test suite. The tool's output comprises the killed and survived mutants, the selected test cases in the test suite, and the mutation score.

Our experimental results in TestSmart indicate that the mutation scores of test suites generated by Manticore vary from 50\% to 100\%. Furthermore, the number of selected test cases can be significantly less than the number of test cases generated by Manticore. It was also observed that the mutation score for test suites generated by symbolic execution (Manticore) depends on the usage of Solidity-specific syntax. Specifically, smart contracts that do not use Ethereum addresses, gas values, transfer function, calling mechanisms, and other similar Solidity-specific functionalities, tend to have higher mutation scores. TestSmart has also evaluated the survived mutants manually. In the following, we discuss the mutants that could not be killed by the test cases generated by symbolic execution and the corresponding reasons.

\begin{enumerate}
\item \textbf{Unintelligent Concretization:} Symbolic execution tools mostly have two possible policies to concretize the values: \textbf{(1)} choosing \textit{random} values from the solved intervals or \textbf{(2)} choosing the \textit{boundary} values. However, our findings show that there are mutants that cannot be killed by any of these policies.

\item \textbf{Need for Intermediate Smart Contracts:} Real-world faults in smart contracts can sometimes only occur after a certain deployment of intermediate smart contracts. The DAO attack is a well-known example of this kind of fault, requiring an attacker's smart contract to exploit the re-entrancy bug through a callback function implementation. Additionally, it is necessary to test the smart contracts that have been published as libraries or interfaces to be able to deploy intermediate smart contracts that have inherited those interfaces, their functions, and the corresponding logic. Unfortunately, symbolic execution tools are unable to generate test data for this class of faults.

\item \textbf{Address Comparison:} In most real-world smart contracts, some condition clauses involve comparisons between the address of the smart contract's caller. In order to obtain higher mutation scores, it is necessary for the test suite to be able to generate transactions with varying addresses. However, symbolic execution tools are not capable of generating such test data, which results in the corresponding mutant not being killed.

\item \textbf{Need for Multi-step Test Cases:} In some cases, errors in smart contracts can only be discovered after several preliminary transactions have been executed. For instance, a fault in the winner calculation of an auction smart contract can only be detected after participants have submitted bids. According to the results obtained from TestSmart, it has been found that symbolic execution tools are unable to produce such test data.
\end{enumerate}

%\subsection{Infection and Reachability Conditions}
 The identification of conditions capable of killing a survived mutant involves the consideration of two pivotal factors:  reachability and infection.  Reachability ensures the execution of the altered line of code by the test case, while infection specifies the conditions under which running the test case can lead to the mutant altering the program's state or execution path compared to the original program.    
 
In order to formulate the reachability condition, we define $\mathcal{C}_{\textit{reach}}$ as the boolean conjunction of all the conditions that must be fulfilled to reach the altered line. To illustrate, consider Listing~\ref{code:reachability-cond} where the mutation modifies line \#3; the reachability condition to reach the altered line can be expressed as ($\alpha \land \beta$).

\noindent\hspace{0.3\textwidth}
\begin{minipage}{.4\textwidth}
\begin{lstlisting}[caption={Reachability Condition},label={code:reachability-cond},mathescape=true]
if ($\alpha$)
    if ($\beta$)
        // altered statement
\end{lstlisting}
\end{minipage}\hfill

\begin{figure*}
	\centering
	\includegraphics[width=0.9\textwidth]{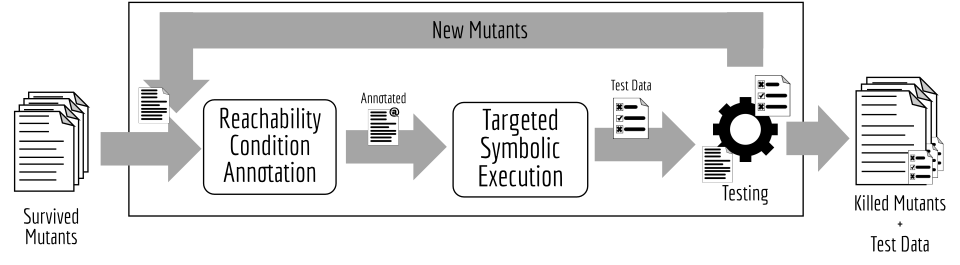}
	\caption{Killing survived mutants using targeted symbolic execution tools}\label{fig:targeted-symbolic-execution-to-kill-mutants}
\end{figure*}

\begin{table}[]
\caption{Categorizing the mutation operators from \cite{andesta2020testing}}\label{tbl:operators}
\begin{tabular}{|p{0.2\linewidth}|p{0.12\linewidth}|p{0.7\linewidth}|}
\hline
\textbf{\specialcell{Operator\\Category}}           & \textbf{\specialcell{\# of \\ Operators}} & \textbf{Description}                                                                                                                                                                                   \\ \hline
Classic Mutation                     & 12                           & The operators that can be applied to any software (including smart contracts), e.g., replacement of arithmetic operators                                                                               \\ \hline
Solidity-Specific Overflow-Underflow & 4                            & Changing the unsigned and signed data types and their sizes. E.g., changing \lstinline|uint16| to \lstinline|uint8|.                                                                                   \\ \hline
\specialcell{Access Control\\ Level}                 & 8                            & Altering the access control for the functions, e.g., changing \lstinline|external| to \lstinline|internal|.                                                                                            \\ \hline
Transaction Call Mechanisms          & 4                            & Mutating the call mechanisms, e.g., altering \lstinline|transfer| to \lstinline|call|.                                                                                                                 \\ \hline
\specialcell{Guard\\Mechanisms}                    & 8                            & Mutating the guard mechanisms and their conditions, e.g., changing the condition inside an \lstinline|assert| to \lstinline|False|. Alternatively, altering \lstinline|assert| to \lstinline|require|. \\ \hline
\specialcell{Transaction\\Origin}                   & 2                            & Mutating \lstinline|msg.sender| and \lstinline|tx.origin|. \\ \hline
\lstinline|selfdestruct|             & 2                            & Removing or changing the line of the \lstinline|selfdestruct| usage.                                                                                                                                   \\ \hline
\specialcell{Constant\\Manipulation}                & 3                            & Manipulating Solidity-specific constants, such as addresses, gas values, and currency-based thresholds.                                                                                                \\ \hline
Manipulating Modifiers               & 2                            & Altering the conditions inside the Solidity modifiers to \lstinline|True| or \lstinline|False|.                                                                                                        \\ \hline
\end{tabular}
\end{table}

Different mutation operators have been proposed for smart contracts in the literature. Considering our proposed mutation operators in~\cite{andesta2020testing}, as described in Table~\ref{tbl:operators}, the infection condition for killing a mutant can be categorized into multiple cases as follow. The infection conditions can be similarly formulated for other proposed mutation operators.

\begin{enumerate}
\item In the scenario where a mutant modifies a condition statement (including \lstinline|if|, \lstinline|while|, assertation, and modifiers statements), it is imperative to ensure that exclusively one of the original conditions or the mutant condition is valid when running under the test case. Let $\mathcal{C}_{\textit{original}}$ represent the original program's condition and $\mathcal{C}_{\textit{mutant}}$ denote the mutated condition. The infection condition can be then formulated as $(\mathcal{C}_{\textit{original}} \oplus \mathcal{C}_{\textit{mutant}})$. For example, when a mutation operator changes the statement \lstinline|assert(a>b)| to \lstinline|assert(a>=b)|, the mutant killing condition is expressed as $(\alpha \land \beta) \land [(a > b) \oplus (a >= b)]$, which is the conjunction of reachability and infection conditions.

\item When a mutant modifies the right-hand side of an assignment statement, it is important to verify that the original and the altered right-hand side are not equal. We can denote the original and altered right-hand side statements as $\mathcal{C}_{\textit{original}}$ and $\mathcal{C}_{\textit{mutant}}$, respectively. The infection condition in this scenario can be expressed as $(\mathcal{C}_{\textit{original}} \neq \mathcal{C}_{\textit{mutant}})$. For instance, the mutant killing condition for the mutant in Listing~\ref{code:reachability-right-hand-side} can be represented as $(\alpha \land \beta) \land [a + b \neq a + 1]$.

\noindent\hspace{0.25\textwidth}
\begin{minipage}{.45\textwidth}
\begin{lstlisting}[caption={Infection Condition for Altering Right-hand Side in Assignement Statements},label={code:reachability-right-hand-side},mathescape=true]
if ($\alpha$)
    if ($\beta$)
        z = a + b // z = a + 1
\end{lstlisting}
\end{minipage}\hfill

\item 
When mutation operators alter access control modifiers, like switching from \lstinline|public| to \lstinline|private|, we define only the reachability condition, not an infection condition. However, to call the function from outside, a smart contract needs to be synthesized using template-based synthesis. A similar process also needs to be repeated for other types of access control modifiers.

\item
For the mutation operators that alter \lstinline|tx.origin| and \lstinline|msg.sender|, the infection condition should be \lstinline|tx.origin| $\neq$ \lstinline|msg.sender|. Also, we should solve a template-based synthesis problem that generates an intermediate smart contract to guarantee the satisfaction of the infection condition.

\item
When mutation testing alters the size of the data types for discovering the underflow and overflow bugs, the infection condition should choose values in the symmetric difference of the data type ranges. For example, when an operator changes \lstinline|uint8| to \lstinline|uint16| for value $x$, the infection condition is $x \in [0, 255] \bigtriangleup [0, 65535]$. Also, any variable usage should be checked as the target line separately for the reachability condition.

\item
On swapping lines mutation operator, firstly, the statements should be converted to symbolic conditions denoted as $\mathcal{C}_{\textit{original}}$ and $\mathcal{C}_{\textit{mutant}}$. Then, the infection condition should ensure that exclusively one of these conditions is satisfied ($\mathcal{C}_{\textit{original}} \oplus \mathcal{C}_{\textit{mutant}}$).

\item
When an operator changes call mechanisms such as ``\lstinline|call|'', ``\lstinline|send|'', and ``\lstinline|transfer|'', it is essential to use the appropriate gas value for the transaction within the infection condition. This is because gas usage differs between these mechanisms. For example, \lstinline|send| only includes a specific amount of gas (represented by 21000 wei), while \lstinline|transfer| sends all the remaining gas on the generated transaction. In addition, it is essential to use template-based synthesis for generating smart contracts to address known bugs such as ``re-entrancy'' as the receiver of the call mechanisms.

\item
When a mutation operator removes or swaps a \lstinline|selfdestruct|, we do not need any infection condition, and only reaching the \lstinline|selfdestruct| line is enough.
\end{enumerate}

Considering the reachability and infection condition for killing a mutant, it can be simplified to the targeted (backward) symbolic execution problem~\cite{li2018targeted,jensen2013automated,arzt2015using} without losing generality.
Moreover, it is important to note that the applications of targeted symbolic execution go beyond generating test cases to kill survived mutants. It can also be used for various purposes, such as detecting false positives in static analysis results, reaching specific lines of code, and identifying input values violating a given safety rule. Hence, this paper aims to provide an approach to generate targeted test data to reach a target line considering a set of given conditions.

%\subsection{Problem Generalization and Formulation}

\noindent In brief, the inputs and outputs of the problem are as follow:

\begin{itemize}
	\item \textbf{Inputs:} A Solidity smart contract, a target line, and an optional safety condition
	\item \textbf{Outputs:} A list of EVM transactions that can reach the target line, while the safety condition is met.
\end{itemize}

Our goal is to propose an algorithm to find Ethereum transaction inputs that can reach a given target line for a given solidity smart contract, while satisfying the given safety conditions.
%\section{High-level Design of Targeted Symbolic Execution for Smart Contract}
\section{\ToolNameNoSpace: A Targeted Symbolic Execution Tool for Smart Contracts}
\label{sec:our-work}

\begin{figure*}
\centering
\includegraphics[width=\textwidth]{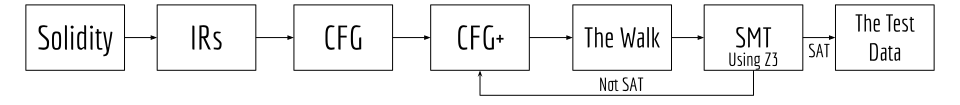}
\caption{Steps for Targeted Symbolic Execution}\label{fig:steps}
\end{figure*}

In this section, we will introduce the high-level design of our proposed framework, which is depicted in Figure~\ref{fig:steps}. Our tool, \ToolNameNoSpace, works by taking a Solidity smart contract, where the target line number along with any safety condition that needs to be satisfied are specified. We use annotations in the smart contract to specify the target line, as well as safety conditions. These conditions should consist of Solidity clauses that can be evaluated to boolean values and only use variables within the current scope.

In the first step, \ToolName converts the smart contract into intermediate representations (IRs).  The IRs can handle all Solidity variables, operators (including assignment, binary, indexing, member, new, push, delete, convert, unpack, array initialization, and call operators), return statements, and conditions. Once \ToolName translates the smart contracts to IRs, it creates the Control Flow Graph (CFG) of the smart contract based on these expressions. In the next step, in addition to other tools that use CFG, \ToolName constructs a graph called CFG+ based on CFG, which includes EVM state (transactions, storage state, and Ethereum balances) along with the control flow. More details will be discussed in Section~\ref{sec:cfg-plus}.

Next, \ToolName attempts to generate a minimal walk from the target to the start node (a node prior to calling the smart contract's constructor) iteratively and verifies its validity (i.e., existence of valid inputs that can generate such a walk) using an SMT solver. If a satisfied walk is found, \ToolName will halt and output the transactions required to reach the target.

\subsection{IR Representation}
As discussed earlier, \ToolName creates IR clauses from the input smart contract as the first step.  As an example, the IR representation for Listing~\ref{code:simple-integer-overflow} is displayed in Listing~\ref{code:ir-simple-integer-overflow}.

\begin{lstlisting}[caption={Simple Interger Overflow Smart Contract~\cite{manticore-sample-codes}},label={code:simple-integer-overflow},numbers=none]
contract Overflow {
    uint16 private sellerBalance=0;

    function add(uint16 value) public {
        sellerBalance += value;
    }
}
\end{lstlisting}

IRs are capable of supporting all Solidity variables, operators, return statements, and conditions. As an example, in Listing~\ref{code:ir-simple-integer-overflow}, line~4 indicates that the function \lstinline|add| adds the input \lstinline|value| to the variable \lstinline|sellerBalance|. Additionally, line~7 initializes the variable \lstinline|sellerBalance| in the constructor of the contract.

\begin{lstlisting}[caption={Simple Interger Overflow Smart Contract~\cite{manticore-sample-codes} in IRs},label={code:ir-simple-integer-overflow}]
Contract Overflow
	Function Overflow.add(uint16) (*)
    	Expression: sellerBalance += value
    	IRs: sellerBalance(uint16) = sellerBalance (c)+ value
	Function Overflow.slitherConstructorVariables() (*)
    	Expression: sellerBalance = 0
    	IRs: sellerBalance(uint16) := 0(uint256)
\end{lstlisting}

Our prototype version utilizes Slither to generate IR (i.e., SlithIR) clauses. Slither~\cite{josselinfeist2018} is a tool that detects over 75 patterns of known bugs and vulnerabilities categorized by their impact and the tool's confidence. Since 2018, this tool has found 40 real-world bugs on smart contracts, preventing losses of over \$67 million~\cite{slither-trophies}. Many other works in the literature use Slither as a foundation for their research. As an example, \cite{zhang2019mpro} uses Slither to compute data dependencies for their symbolic execution tool to improve scalability. Additionally, \cite{yu2020smart} provides code repair recommendations for bugs detected by Slither.

\subsection{CFG and CFG+}\label{sec:cfg-plus}
Considering the nature of smart contracts, we could not use CFG for the purpose of targeted symbolic execution. The reasons are the following:
\begin{itemize}
	\item Unlike traditional software, the CFG of smart contracts has multiple entry points, since users can call any contract's public function.
	\item We need to store the history of previous transactions and the storage leading up to the construction of the smart contract on the EVM.
\end{itemize}
To overcome the mentioned challenges, we introduce CFG+ for smart contracts. Before discussing CFG+, we first present the formal definition of CFG for smart contracts (based on the Ethereum Yellow Paper~\cite{wood2014ethereum}).

\begin{definition}[Control Flow Graph (CFG)]
The control flow graph CFG for smart contracts is a multi-graph that includes two directed graphs ${\text{CFG}}_{\textit{constructors}}$ = $(V, E, S, T)$ and  ${\text{CFG}}_{\textit{functions}}$ = $(V', E', S', T')$ comprising:
\begin{itemize}
\item $(V, E)$ and $(V', E')$ are CFG graphs for the constructor and other functions in the smart contract, respectively, similar to other non-smart-contract software. In these graphs, program statements are nodes, where an edge $(v_1, v_2)$ indicates that $v_2$ can follow $v_1$ during execution.
\item The initial states for these graphs are $S \subset V$ and $S' \subset V'$. It is worth mentioning that for each node $v \in V$ and $v' \in V'$, there exists a node $s \in S$ and $s' \in S'$ respectively, such that a path can be found from $s$ to $v$ and $s'$ to $v'$.

\item $T \subset V$ and $T' \subset V'$ are the final states.
\end{itemize}
\end{definition}

Our enhanced version of CFG (CFG+) is defined as follows, according to the formal definition of the CFG for smart contracts.

\begin{definition}[CFG+]
Considering the CFG including ${\text{CFG}}_{\textit{constructors}}$ = $(V, E, S, T)$ and  ${\text{CFG}}_{\textit{functions}}$ = $(V', E', S', T')$ the directed graph $\text{CFG+} = (V'', E'', S'', T'')$  is defined as follows:

\begin{itemize}
	\item $V'' = V \cup V' \cup \{\textit{``start"}, \textit{``end"}, \textit{``constructed (active)"}, \textit{``TX processed"} \}$ is the list of the nodes in CFG+ which includes four auxilary nodes.
	\item The edges for CFG+ are defined as:
	\begin{align*}
		E'' &= E \cup E' \cup \{(\textit{``start"}, s) | \forall s \in S \} \cup \{(t, \textit{``constructed (active)"}) | \forall t \in T \} \\
			&\cup \{(\textit{``constructed (active)"}, s') | \forall s' \in S' \} \cup \{(t', \textit{``TX processed"}) | \forall t' \in T' \} \\
			&\cup \{(\textit{``TX processed"}, \textit{``constructed (active)"}), (\textit{``TX processed"}, \textit{``end"})\} 
	\end{align*}
	Which tries to connect the auxiliary nodes to the subgraphs of ${\text{CFG}}_{\textit{constructors}}$ and ${\text{CFG}}_{\textit{functions}}$.
	\item $S'' = \{``start"\}$ is the initial node and $T'' = \{``end"\}$ is the final node.
\end{itemize}
\end{definition}

\begin{figure*}
	\centering
	\includegraphics[width=\textwidth]{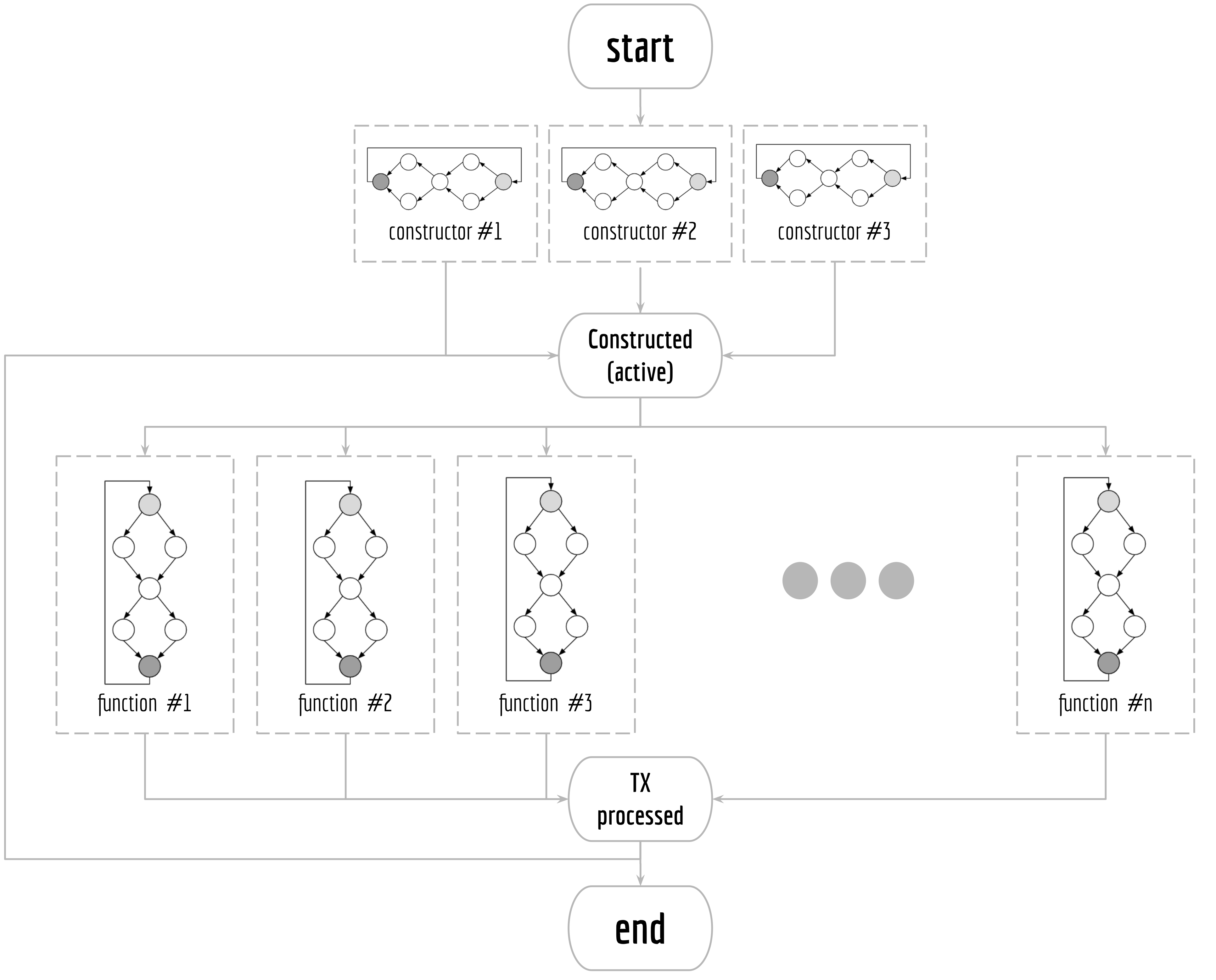}
	\caption{CFG+: The enhanced CFG that considers the EVM storage state}\label{fig:cfg-plus}
\end{figure*}

Intuitively, CFG+ is an upgraded version of the control flow graph (CFG) that has the ability to retain the history of past transactions and their results. The diagram in Figure~\ref{fig:cfg-plus} demonstrates that CFG+ begins with a new EVM with no deployed smart contracts but some predefined Ethereum accounts (i.e. ``start"). From there, any smart contract constructors can be executed, which changes the smart contract's state to ``constructed (active)." After that, any user (or other smart contracts) can call any of the smart contract functions and move the CFG+ to the ``TX processed" state. During the transaction execution, the EVM may either reach the target and terminate, or the test data may decide to initiate another transaction (i.e., returning to ``constructed (active)" once again).

It is important to note that global variables are shared during the CFG+ traversal and are reset when the ``start node" is reached. However, local variables are reset when new transactions are started. Additionally, when each transaction is started, \ToolName resets the transaction-related variables such as \lstinline|msg.sender|, \lstinline|msg.value|, \lstinline|tx.origin|, timestamps, user balances, etc., on the "constructed (active)" state.

%Please note that in Figure~\ref{fig:cfg-plus}, we have removed private functions, functions called from libraries, and methods called from other smart contracts to improve the readability and simplicity of the diagram. However, these can still be modeled within the ``functions" boxes and will not reach any of the ``constructed (active)" or ``TX processed" nodes.\FF{What do you mean?}

\begin{figure}
\centering
\includegraphics[width=0.85\textwidth]{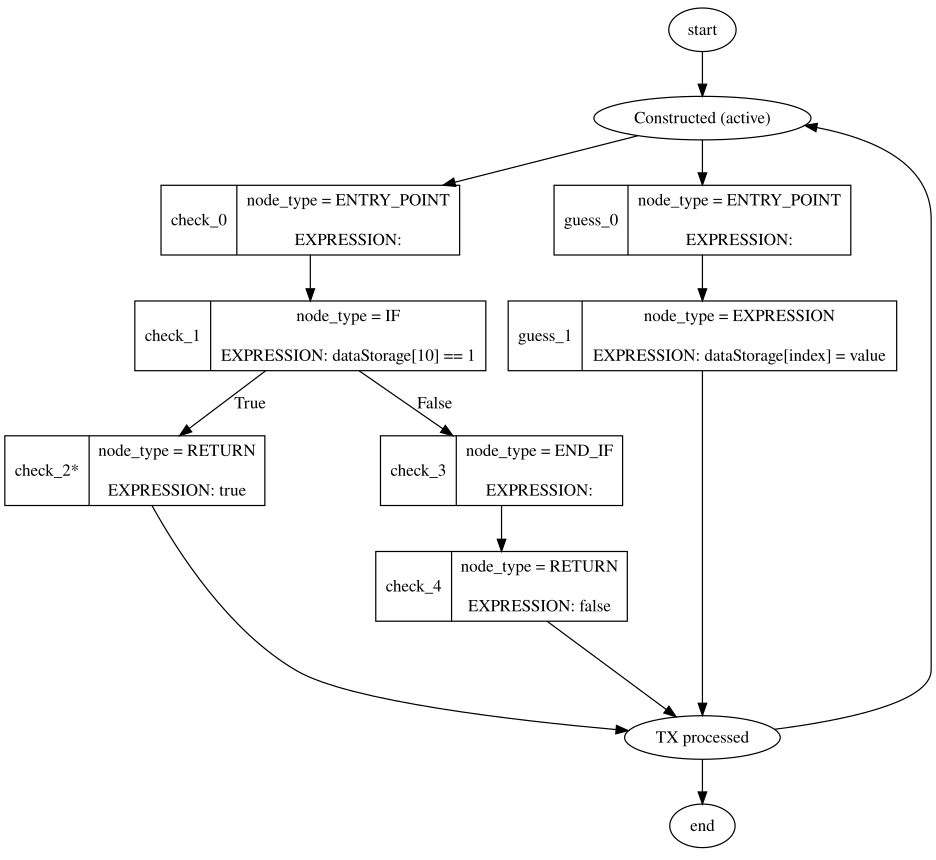}
\caption{CFG+ for Simple Guess\&Check sample using mapping}\label{fig:simple-mapping-cfg-plus}
\end{figure}

To generate the CFG+, \ToolName utilizes Slither for the creation of the CFG of the smart contracts. Then, it forms the constructors and other functions. A sample of CFG+ is illustrated in Figure~\ref{fig:simple-mapping-cfg-plus} related to the smart contract in Listing~\ref{code:simple-mapping}. This smart contract has a mapping named ``dataStorage" and two functions named ``guess" and ``check". An Ethereum account can guess an index and a value and the winner is the user who chooses index 10 and value 1.

\begin{lstlisting}[caption={Simple Guess\&Check sample using mappings},label={code:simple-mapping},numbers=none]
contract mappingSample {
    mapping(uint => uint) public dataStorage;

    function check() public returns (bool) {
        if (dataStorage[10] == 1)
            return true;  // @target
        return false;
    }

    function guess(uint index, uint value) public {
        dataStorage[index] = value;
    }
}
\end{lstlisting}

Once CFG+ is established, \ToolName initiates the traversal algorithm for the minimal satisfiable walk. However, to simplify the process, the algorithm utilizes the reversed version of CFG+, which is mathematically the transposed graph.

\subsection{Discovering Minimal Satisfiable Walk}\label{sec:minimal-walk}
Based on reversed CFG+, \ToolName tries to find test data, including Ethereum transactions that reach the target, i.e., \textit{minimal walk}. This section will cover the \ToolName algorithm for discovering this test data. To begin, let us establish the definition of the minimal walk.

\begin{definition}[Minimal Walk]
	The ``Minimal Walk" $\mathcal{W}$ refers to the optimum path on the reversed CFG+ graph from the target node $t$ to the starting node $s$, as determined by the "Heuristic" $\mathcal{H}$.
\end{definition}

Based on the definition of the minimal walk, we can now define ``Minimal Satisfiable Walk."

\begin{definition}[Minimal Satisifiable Walk]
	The ``Minimal Satisifiable Walk" refers to the minimal walk $\mathcal{W}$ that satisfies the SMT problem based on the IRs on CFG+ during the walk and also satisfies the given safety condition.
\end{definition}

According to these definitions, \ToolName aims to identify the minimal satisfiable walk from the target node to the starting node while considering the tester's target condition. To accomplish this, \ToolName allows for the use of any heuristic that can be customized. Algorithm~\ref{alg:minimal-sat-walk} demonstrates how \ToolName traverses CFG+ based on the given heuristic. This algorithm works in increments, starting from the target node and searching for the best neighbor to extend each time. If the chosen neighbor is the CFG+ start node and the SMT problem is SAT, the algorithm stops and returns the test data. However, if the path from the target node to the selected neighbor is not satisfiable, it removes this walk and looks for other nodes to extend. If the increment is SAT (but not the start node), it is added as another frontier to expand later based on the heuristic.

\begin{algorithm}
	\caption{Minimal Satisfiable Walk Algorithm}\label{alg:minimal-sat-walk}
	\begin{algorithmic}
		\Function{walk}{\textit{targetNode}}
		\State \textit{tree: Tree} $= \{targetNode, \}$
		\State \quad
		
		\While{\textit{tree.hasExtendableLeafs()}}		
		\State \textit{options = tree.leafsNeigbours()}
		\State \textit{bestOption} = $\operatorname*{\textit{arg\,min}}_{\textit{option} \in \textit{options}} \mathcal{H}(\textit{tree, option})$
		\State \quad
		
		\If{$\textit{bestOption} == \textit{``startNode"} \And \textit{SAT}$}
		\State \Return \textit{tree.extend(bestOption)}
		\EndIf
		\State \quad
		
		\If{\textit{bestOption is not SAT}}
		\State \textit{tree.doNotContinue(bestOption)}
		\Else
		\State \textit{tree.extend(bestOption)}
		\EndIf
		\EndWhile
		\EndFunction
	\end{algorithmic}
\end{algorithm}

It should be noted that while \ToolName is capable of accepting any heuristics for traversing walks on CFG+ to find the minimal satisfiable walk, we have implemented a naive heuristic based on the Floyd-Warshall algorithm as the default heuristic in the prototype version. We also proposed another heuristic based on the define-usage of the state variables during the smart contract execution. This heuristic determines if a state variable in the current tree (based on the Algorithm~\ref{alg:minimal-sat-walk}) is being read but not written. If such a variable exists, the heuristic will only choose options that can write these variables and use Floyd-Warshall to determine the best option from multiple options. The heuristic will return infinity for options that cannot write the required state variables.

\subsection{Solving SMT and Concritization}\label{sec:smt-and-concritization}
As we have discussed earlier, the Minimal Satisfiable Walk Algorithm requires \ToolName to transform walks into SAT/SMT problems and use a solver to verify if they are satisfiable. Once the solver confirms that the walk can be satisfiable and reaches the entry point, \ToolName should attempt to transform symbols into absolute values. In this section, we will explain this process in detail, and interested readers can refer to our prototype version on GitHub for implementation details.

\subsubsection{Static Single Assignment Form (SSA)}
The IR expressions must be written in Static Single Assignment form, which means a variable should not be assigned twice during the SAT/SMT clauses. Hence, \ToolName generates SSA clauses based on IR grammar rules that include variable assignments. Consider the statement $var = 10; var = var + 20;$. If we transform these statements to $var = 10 \land var = var + 20$, this set would be unsatisfiable. Therefore, \ToolName generates indexed variables on each variable assignment. The statement will be transformed to $var_1 = 10 \land var_2 = var_1 + 20$, which will be satisfiable. It is also worth noting that exploring the statements backward makes converting expressions to SSA in \ToolName more complicated. According to the SlithIR~\cite{feist2019slither} grammar, the most important grammar that should be considered in the Static Single Assignment (SSA) form are as follows:
\begin{itemize}
	\item Assignment
	\item Binary/Unary Operations
	\item Expressions with Indexing Operator
	\item Conditions
	\item Initializing Functions
	\item Processing RValues
\end{itemize}

\subsubsection{Complex Data Structures}
As the SAT/SMT solvers only support primitive data types, the Solidity complex data types should be converted to these primitives. An uninterpreted SMT function is declared for ``mappings" and ``arrays" with the ``key" and ``Index" as an input parameter. Also, we should add an auxiliary index parameter for the process of making these expressions a static single assignment. As an example, assuming an array of integers \lstinline|arr|, an uninterpreted SMT function $\textit{arr}_{i}(\textit{indx})$ should be declared with the index $\textit{indx}$, SSA index $i$ as its parameters. Then, for the new assignment of value $v$ to numeric index $\textit{indx}$ in array $arr$, with current SSA index $i$ and resulting new SSA index $j$, the following SMT clauses are applicable:
$$
			\textit{arr}_{j}(\textit{indx}) = v	\land \forall_{k \neq \textit{indx}} \textit{arr}_{j}(\textit{k}) = \textit{arr}_{i}(\textit{k})
$$

Additionally, when dealing with structs and events, \ToolName treats every member as an independent variable in the SAT/SMT clauses. The same approach is taken for EVM and transaction-related variables. For example, an independent variable is defined for each member of the special variables like \lstinline|tx|, \lstinline|msg|, and \lstinline|block|. Interested readers can refer to our prototype version on GitHub for implementation details.

\subsubsection{Concritization and Test Data Generation}
After identifying the minimal satisfiable walk, \ToolName utilizes the SAT/SMT solver to determine absolute values for the variables and then converts them into EVM transactions. Each transaction includes the function called, the Ethereum account initiating the call, input parameters for the function, gas value, and assigned ETH.

\subsubsection{The Prototype Version}
We have completely implemented \ToolNameNoSpace, and made it publicly available for interested users  on GitHub\footnote{https://github.com/professormahi/griffin-targeted-symbolic-execution}. The prototype version of \ToolName tool uses the CFG+ to traverse all potential walks from the target node to the start node and uses a heuristic to find the minimal walk. To determine if a walk is satisfiable (i.e., has equivalent test data), \ToolName converts IRs to SMT clauses and employs an SMT solver like Z3 to solve the SMT problem.
To create the SMT clauses, we designed the semantics of SlithIRs grammar for translating IRs to SMT clauses and implemented the translator using PLY~\cite{ply}. \ToolName uses, but is not limited to, the Z3 theorem prover~\cite{z3} to solve the SMT problem and concretization process.

\section{Evaluation}\label{sec:evaluation}

 To evaluate \ToolNameNoSpace's efficiency, we collected a comprehensive set of smart contracts with annotated targets and target conditions. We selected these sample smart contracts from various sources, including samples from other smart contract testing tools and prominent real-world bugs, Solidity samples, the test suite for Manticore symbolic execution tools, and our previous work~\cite{Goodin2021,manticore-sample-codes,fooladgar2021testsmart}. The source code for these samples is available for interested users on Github\footnote{https://github.com/professormahi/griffin-targeted-symbolic-execution}.

We conducted our evaluation using a personal laptop equipped with an Intel Corei7-1255U CPU and 32GB of memory, and hence, most Solidity developers and testers can reproduce similar results. We tested \ToolName on 12 smart contracts with varying software size (LoC) and complexity (number of nodes in CFG), as shown in Table~\ref{tbl:evaluation-results}. Each smart contract was tested 10 times, and the average execution time of our tool \ToolName is reported in Table~\ref{tbl:evaluation-results}.

\begin{table}[tbh]
\caption{Evaluation Results}\label{tbl:evaluation-results}
\resizebox{\textwidth}{!}{
\begin{tabular}{|l|c|c|c|c|}
\hline
\textbf{Smart Contract}                      & \textbf{\#LoC} & \textbf{\# Generated Walks} & \textbf{\# CFG Nodes} & \textbf{Execution Time (s)} \\ \hline
Mutant Killing Sample                        & 11             & 6                           & 12                    & 0.08                        \\ \hline
Simple Mapping [\lstinline|address => uint|] & 18             & 11                          & 15                    & 0.40                        \\ \hline
Two TX Overflow [Static Analysis]                             & 24             & 19                          & 20                    & 0.27                         \\ \hline
Simple Condition Checking                    & 8              & 6                           & 9                     & 0.07                        \\ \hline
Simple \lstinline|msg.value| Check           & 9              & 5                           & 9                     & 0.07                        \\ \hline
Simple Integer Overflow                      & 11             & 15                          & 10                    & 0.21                        \\ \hline
Mapping Sample [\lstinline|uint => uint|]    & 12             & 10                          & 11                    & 0.14                        \\ \hline
Dummy MonoX Bug                              & 21             & 13                          & 17                    & 0.34                         \\ \hline
Multi TX Auction (5 TXs)                             & 19             & 884                         & 20                    & 23.70                       \\ \hline
Internal Call Sample                         & 21             & 15                         & 20                    & 0.15                       \\ \hline
SafeMath Sample                              & 46             & 13                         & 17                    & 0.17                       \\ \hline
ERC20 Token                           & 118            & 28                         & 88                    & 3.60                       \\ \hline
\end{tabular}
}
\end{table}

The results displayed in Table~\ref{tbl:evaluation-results} indicate that the execution time of \ToolName is not significantly correlated with either the size of the software (measured in lines of code) or its complexity (measured by the number of CFG nodes). However, it is strongly linked to the number of paths that must be explored to locate the desired test data. This means the execution time is linked to the logic of the smart contract and the test data, not to the size of the smart contract. Therefore, to achieve optimal performance, developers or testers should select or create a suitable heuristic based on their specific software or target line.

Consider the smart contract ``Multi TX" (demonstrated on Listing~\ref{code:multi-tx} based on our previous work~\cite{fooladgar2021testsmart}). To find the minimal satisfiable walk and generate the required test data, \ToolName must explore 884 walks. The ``Multi TX" is a simple auction smart contract that can have up to five bid thresholds (configurable during smart contract deployment). After that, the highest bid becomes the winner. \ToolName may generate several paths that call the function \lstinline|check|, which is unnecessary. To reach the target line, \ToolName must generate test data with 5 transactions calling the \lstinline|bid| function. However, we can significantly improve performance if we use a heuristic based on \textit{state variables usage} which described in Section~\ref{sec:minimal-walk} instead of the Floyd-Warshall algorithm.

\begin{lstlisting}[caption={Multi TX Smart Contract},label={code:multi-tx}]
contract multi_tx {
    uint public counter = 0;
    uint public maximum_bid = 0;
    uint private threshold = 5;

    function bid(uint value) public {
        counter += 1;
        if (value > maximum_bid)
            maximum_bid = value;
    }

    function check() public returns (uint) {
        if (counter == threshold)
            return maximum_bid;  // @target maximum_bid > 100
        return 0;
    }
}
\end{lstlisting}

\begin{figure*}
\centering
\includegraphics[width=0.8\textwidth]{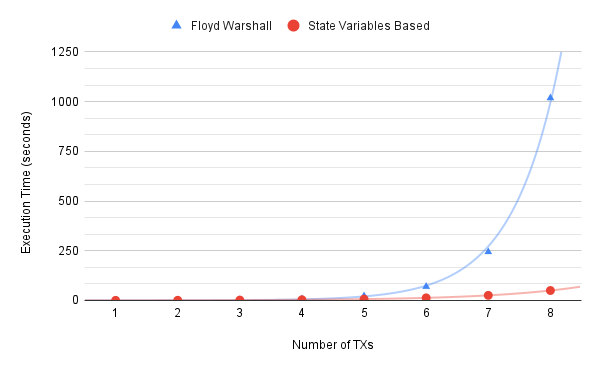}
\caption{Comparing the execution time of Floyd-Warshall Heuristic and State Variables Based for different thresholds in Listings~\ref{code:multi-tx}}\label{fig:heuristic-evaluation}
\end{figure*}

In Figure~\ref{fig:heuristic-evaluation}, a comparison is made between the Floyd-Warshall and state-variables-based heuristic. The results show that the state-variables-based heuristic has a significantly lower growth rate and a more reasonable time than the Floyd-Warshall heuristic. This is based on the number of transactions that need to be generated for the smart contracts in Listings~\ref{code:multi-tx}. These results demonstrate how choosing an effective heuristic can reduce the number of generated paths during the CFG+ traversal, subsequently reducing execution time. We cannot make a general decision about which heuristic has better performance, as their performance may vary across different smart contracts and target lines. Furthermore, developers have the option to implement custom heuristics in \ToolName to achieve improved performance. Also, Researchers interested in comparing their proposed heuristics with others can use \ToolName and provide discussions on the conditions under which their heuristic demonstrates better performance.

In our previous study, we make use of TestSmart~\cite{fooladgar2021testsmart} to verify whether Manticore, a symbolic execution tool, is capable of killing all mutants for a smart contract. These experiments indicate that symbolic execution tools suffer from unintelligent concretization and cannot generate multi-transaction test data using Manticore. 
In these experiments, we utilize the ``Mutant Killing Sample" to check the reachability condition on a mutant. This condition is a combination of the negation of the original condition and the mutant condition for mutations on the conditions. Additionally, as previously discussed, \ToolName is able to generate test data on ``Multi TX" within a reasonable amount of time.

It is worth noting that \ToolName could also have prevented a bug that caused a 31 million-dollar loss on MonoX~\cite{Goodin2021}. In the MonoX attack, the attacker swapped a token with itself, resulting in the decrease in the prices of the input token being overwritten by the increase in the output token's price, as they were the same. This attack led to an increase in the token's price. The tester could have used \ToolName with a simple target condition that checks if the input token's price decreases and the output token's price increases in all cases to generate the test data for this bug. This experiment has demonstrated that even though the standard static analysis and dynamic testing tools may not be able to detect a bug caused by a logical fault, utilizing \ToolName with appropriate safety measures can generate test data to help identify such bugs.

Moreover, as we discussed earlier, \ToolName can be used to generate test data for static analysis tool alarms. For example, ``Two TX Overflow" is a smart contract marked as a possible overflow by Manticore's internal static analyzer. Our tool shows that the overflow fault will only be revealed if there are two separate Ethereum transactions. The first transaction should initialize the smart contract values, and the second transaction can trigger an overflow in the statement.
\section{Related Works}\label{sec:related-works}
%\begin{itemize}
%	\item Symbolic Executions tools
%	\begin{itemize}
%		\item Manticore
%		\item Mythril
%		\item Oyente
%		\item TestSmart (our work)
%	\end{itemize}
%
%	\item Automated test data generation using search algorithms
%	\begin{itemize}

%\end{itemize}
%		\item tools using Genetic Algorithms, etc.
%	\end{itemize}
%	
%	\item Other targeted test data generation there is any
%\end{itemize}

There are several works on the analysis of smart contracts. We briefly discuss them in two categories; static analysis, and dynamic testing.

%\subsection{Formal Verification}
%FSolidM, as described in \cite{mavridou2018designing}, is a tool that takes a finite state machine representing a smart contract's design as input. It then uses model-checking techniques to verify formal LTL specifications. If the verification is successful, the tool generates the source code of the smart contract based on the given model as an output. 
%It is important to note that FSolidM comes with pre-set format LTL specifications to detect the most commonly repeated bug patterns. Additionally, it can model-check any custom specification.

\subsection{Static Analysis}
Many researchers have studied the static analysis of Solidity smart contracts.
Slither~\cite{josselinfeist2018} is a tool that detects over 75 patterns of known bugs and vulnerabilities categorized by their impact and the tool's confidence. Since 2018, this tool has found 40 real-world bugs on smart contracts, preventing losses of over \$67 million~\cite{slither-trophies}. Many other works in the literature use Slither as a foundation for their research. As an example, \cite{zhang2019mpro} uses Slither to compute data dependencies for their symbolic execution tool to improve scalability. Additionally, \cite{yu2020smart} provides code repair recommendations for bugs detected by Slither.

SmartCheck~\cite{tikhomirov2018smartcheck} converts Solidity smart contracts to an abstract syntax tree in XML format. The tool then uses XML queries to identify known bug patterns. Some IDE integration static analysis tools, such as one inside the Remix IDE, can find bad smells and known bug patterns on the smart contracts inside the IDE~\cite{remix-ide-static-analysis}. 

 Static analysis techniques, while valuable in bug detection, do have their limitations, such as the following:
\begin{enumerate}
	\item Detection of Known Bug Patterns: Static analysis relies on predefined rules and patterns to identify potential issues in code. This means that it is effective in detecting known bug patterns for which rules have been explicitly defined. However, it may struggle with identifying novel or uncommon bugs that do not fit into the established patterns. In contrast, dynamic analysis techniques, such as testing and runtime analysis, may have a better chance of uncovering unforeseen issues during the execution of the program.
	\item False Positives and Unsound Analysis:
	Static analysis tools can produce false positive results, meaning they may flag code as potentially problematic when, in reality, there is no bug. This can happen due to the complexity of code or limitations in the analysis algorithms. False positives can lead to wasted time and effort if developers investigate and fix non-existent issues.
\end{enumerate}

\subsection{Dynamic Testing}
The best analysis practice is to use a combination of static and dynamic analysis techniques to maximize bug detection capabilities. Static analysis can be employed early in the development process to catch issues before the code is executed, while dynamic analysis can be used to find runtime-specific problems that may not be apparent through static examination alone. Therefore, it is important to use dynamic testing techniques to analyze smart contracts as a complementary method.

Truffle~\cite{truffle-website} and Waffle~\cite{waffle-website} are testing frameworks developed for Solidity smart contract developers. Truffle is a comprehensive test suite that provides developers with a test framework, test blockchain network (i.e., Ganache), and mocking framework. Waffle also provides a similar toolchain, which is claimed to be easier to use. It's worth noting that Truffle and Ganache have recently been discontinued. Hardhat~\cite{hardhat} is another popular Ethereum development environment. It is built with JavaScript and maintained by Nomiclabs. Hardhat functions as an extensible framework, providing comprehensive tools for managing the entire smart contract lifecycle. These tools include compiling, deploying, testing, and debugging your contracts. 

Oyente~\cite{luu2016making} is a symbolic execution tool that can be run on EVM byte-code version of smart contract and tries to find known bug patterns such as timestamp dependency, re-entrancy, etc. Luu et al. evaluated their tool on more than 19000 Ethereum smart contracts in this work and found about 9000 known bugs and vulnerability patterns.
Mythril~\cite{mythril-github} is another symbolic execution tool for Solidity smart contracts, capable of detecting various vulnerabilities and bugs such as safety failures, overflow and underflows, unauthorized access to EVM storage, and bad smells on delegated calls. Harvey~\cite{Wustholz}, another tool from the Mythril developers, is a greybox fuzzer for Solidity smart contracts. In contrast to similar tools, this tool is able to generate multi-transaction test data to reach bugs and vulnerabilities.
Manticore~\cite{mossberg2019manticore} is a dynamic symbolic execution tool that is open-source and supports various formats, including EVM, Linux ELF binaries, and WASM modules. It emulates the Ethereum virtual machine symbolically, which enables it to represent transactions, storage, global values, and EVM's current state. Developers can interact with Manticore and even integrate it into their toolchain using the APIs provided.
SmarTest is also a symbolic execution tool that uses machine learning to learn the language model and a language-model-guided symbolic execution approach to generate a test suite of vulnerable transaction sequences~\cite{So2021}.

There is also another group of dynamic testing tools known as search-based automated test-data generation tools. One example of such a tool is AGSolT~\cite{Driessen2021}, which utilizes genetic algorithms and fuzzing to generate test data. In their research, Driessen et al. implemented a fitness function known as ``normalized branch distance'' to determine which branch is closer to being covered, resulting in higher branch coverage. Syntest~\cite{olsthoorn2022syntest} also uses genetic algorithms and random testing to achieve over 70\% branch and statement coverage on more than 20 smart contracts.

%\subsubsection{Mutation Testing}
%Mutation testing is a powerful technique for white-box testing. It involves making small syntactical changes to the source code, referred to as mutation operators. The concept behind this technique is that faults in software written by experienced developers are often due to tiny errors in syntax. For instance, a developer might accidentally use \lstinline|>| instead of \lstinline|>=|. To implement mutation testing, mutation operators are applied to the source code of the target program one at a time to create mutants. These mutants are then evaluated along with the original program by running a set of test cases (test suite). If the output of the program and a mutant differ on a test case, the mutant is killed, and the test case is selected. The chosen test cases can identify the difference between the program and its mutants, making them effective in identifying bugs.

In the literature, there have been numerous proposals for mutation testing in smart contracts. One is SuMo~\cite{barboni2021sumo}, a tool developed by Barboni et al. that implements 44 mutation operators based on the latest Solidity documentation. The authors evaluated the effectiveness of these operators using two real-world smart contracts and associated test suites. The experimental results indicate that the mutation score for both contracts is moderately low. In another study, Chapman et al. suggested 61 mutation operators and tested them on three contracts and their corresponding test suites~\cite{chapman2019deviant}. Despite having high branch or statement coverage, the authors found that even well-designed test suites may result in low mutation scores. Also, some other similar works are trying to propose mutation operators such as \cite{honig2019practical,li2019musc} or even scalability of mutation testing in smart contracts \cite{hartel2020mutation}.

In~\cite{andesta2020testing}, we conducted an extensive study on known bugs in smart contracts and proposed 49 Solidity-specific mutation operators across 10 classes based on these faults. In this study, we found that these operators could replicate 10 out of 15 of the most expensive real-world Solidity bugs. Later, we developed TestSmart~\cite{fooladgar2021testsmart}, which aims to fill the gap between generating test suites and selecting test cases through mutation testing. TestSmart utilizes Manticore for generating test data and mutation operators introduced in~\cite{andesta2020testing} to evaluate the effectiveness of symbolic execution tools such as Manticore. Our findings indicate that symbolic execution tools fall short in concretization, are unable to produce intermediate smart contracts necessary for test cases, cannot generate test cases that require multiple transactions, and fail to kill address comparison mutants. In this paper, we tried to tackle these challenges using targeted symbolic execution.
\section{Conclusion and Future Works}\label{sec:conclusion}
In this paper, we presented  the first targeted symbolic execution tool for Solidity smart contracts, called \ToolNameNoSpace.  Our tool generates test data to reach a smart contract's target line, considering a safety condition. Additionally, we discuss how our framework can verify alarms generated by static analysis tools, produce test data for the reachability/infection condition of killing mutants, and generate test data for manual target annotation of testers and developers.

Our study demonstrated that our tool efficiently detects the most commonly occurring real-world bugs on smart contracts and other testing tools in a reasonable amount of time. Additionally, we discussed the possibility of using customized heuristics in our framework to expand the paths from entry points to the target line. Our results indicated that selecting appropriate heuristics significantly reduces the execution time. 

During this paper, we presented two heuristics in our prototype implementation. The first heuristic was based on the Floyd-Warshall algorithm, which selects the shortest walk first. However, it was observed that this heuristic had a high execution time on targets that require multiple transactions. Hence, we propose a heuristic based on state variables. It discovers walks to write state variables that were read but not written yet while exploring reversed CFG+. Our results showed that the execution time significantly decreases by using this heuristic.

In our future work, we aim to develop additional heuristics and determine which ones best suit specific target lines and conditions. Additionally, we plan to enhance our prototype version by incorporating a wider range of SMT-solving tools based on the user's preferences.

\bibliographystyle{elsarticle-num}
\bibliography{ref}

\begin{thebibliography}{10}
\expandafter\ifx\csname url\endcsname\relax
  \def\url#1{\texttt{#1}}\fi
\expandafter\ifx\csname urlprefix\endcsname\relax\def\urlprefix{URL }\fi
\expandafter\ifx\csname href\endcsname\relax
  \def\href#1#2{#2} \def\path#1{#1}\fi

\bibitem{theDAO}
T.~Stephan, {A Primer to Decentralized Autonomous Organizations (DAOs)},
  \url{https://medium.com/ursium-blog/a-primer-to-the-decentralized-autonomous-organization-dao-69fb125bd3cd},
  March 3th, 2016 (2016).

\bibitem{Goodin2021}
D.~Goodin, {Really stupid “smart contract” bug let hackers steal \$31
  million in digital coin},
  \url{https://arstechnica.com/information-technology/2021/12/hackers-drain-31-million-from-cryptocurrency-service-monox-finance/}
  (Feb 2021).

\bibitem{mavridou2018tool}
A.~Mavridou, A.~Laszka, Tool demonstration: Fsolidm for designing secure
  ethereum smart contracts, in: International conference on principles of
  security and trust, Springer, 2018, pp. 270--277.

\bibitem{mavridou2018designing}
A.~Mavridou, A.~Laszka, Designing secure ethereum smart contracts: A finite
  state machine based approach, in: International Conference on Financial
  Cryptography and Data Security, Springer, 2018, pp. 523--540.

\bibitem{josselinfeist2018}
josselinfeist, Slither – a solidity static analysis framework | trail of bits
  blog,
  \url{https://blog.trailofbits.com/2018/10/19/slither-a-solidity-static-analysis-framework/}
  (2018).

\bibitem{feist2019slither}
J.~Feist, G.~Grieco, A.~Groce, Slither: a static analysis framework for smart
  contracts, in: 2019 IEEE/ACM 2nd International Workshop on Emerging Trends in
  Software Engineering for Blockchain (WETSEB), IEEE, 2019, pp. 8--15.

\bibitem{tikhomirov2018smartcheck}
S.~Tikhomirov, E.~Voskresenskaya, I.~Ivanitskiy, R.~Takhaviev, E.~Marchenko,
  Y.~Alexandrov, Smartcheck: Static analysis of ethereum smart contracts, in:
  Proceedings of the 1st International Workshop on Emerging Trends in Software
  Engineering for Blockchain, 2018, pp. 9--16.

\bibitem{remix-ide-static-analysis}
RemixIDE, Solidity static analysis — remix ide,
  \url{https://remix-ide.readthedocs.io/en/latest/static_analysis.html} (2023).

\bibitem{truffle-website}
T.~Team, Truffle suite, \url{https://trufflesuite.com/} (2023).

\bibitem{waffle-website}
W.~Team, Waffle, \url{https://getwaffle.io/} (2023).

\bibitem{ellul2018runtime}
J.~Ellul, G.~J. Pace, Runtime verification of ethereum smart contracts, in:
  14th European Dependable Computing Conference (EDCC), IEEE, 2018, pp.
  158--163.

\bibitem{Wustholz}
V.~W{\"u}stholz, M.~Christakis, Harvey: A greybox fuzzer for smart contracts,
  in: Proceedings of the 28th ACM Joint Meeting on European Software
  Engineering Conference and Symposium on the Foundations of Software
  Engineering, 2020, pp. 1398--1409.

\bibitem{luu2016making}
L.~Luu, D.-H. Chu, H.~Olickel, P.~Saxena, A.~Hobor, Making smart contracts
  smarter, in: Proceedings of the ACM SIGSAC conference on computer and
  communications security, ACM, 2016, pp. 254--269.

\bibitem{mythril-github}
Consensys, \href{https://github.com/ConsenSys/mythril}{Consensys/mythril:
  Security analysis tool for evm bytecode. supports smart contracts built for
  ethereum, hedera, quorum, vechain, roostock, tron and other evm-compatible
  blockchains.} (2023).
\newline\urlprefix\url{https://github.com/ConsenSys/mythril}

\bibitem{mossberg2019manticore}
M.~Mossberg, F.~Manzano, E.~Hennenfent, A.~Groce, G.~Grieco, J.~Feist,
  T.~Brunson, A.~Dinaburg, Manticore: A user-friendly symbolic execution
  framework for binaries and smart contracts, in: 2019 34th IEEE/ACM
  International Conference on Automated Software Engineering (ASE), IEEE, 2019,
  pp. 1186--1189.

\bibitem{So2021}
S.~So, S.~Hong, H.~Oh, $\{$SmarTest$\}$: Effectively hunting vulnerable
  transaction sequences in smart contracts through language
  $\{$Model-Guided$\}$ symbolic execution (2021).

\bibitem{fooladgar2021testsmart}
M.~Fooladgar, A.~Arefzadeh, F.~Faghih, Testsmart: A tool for automated
  generation of effective test cases for smart contracts, in: 2021 11th
  International Conference on Computer Engineering and Knowledge (ICCKE), IEEE,
  2021, pp. 476--481.

\bibitem{chaliasos2024smart}
S.~Chaliasos, M.~A. Charalambous, L.~Zhou, R.~Galanopoulou, A.~Gervais,
  D.~Mitropoulos, B.~Livshits, Smart contract and defi security tools: Do they
  meet the needs of practitioners?, in: Proceedings of the 46th IEEE/ACM
  International Conference on Software Engineering, 2024, pp. 1--13.

\bibitem{nikolic2018finding}
I.~Nikoli{\'c}, A.~Kolluri, I.~Sergey, P.~Saxena, A.~Hobor, Finding the greedy,
  prodigal, and suicidal contracts at scale, in: Proceedings of the 34th Annual
  Computer Security Applications Conference, ACM, 2018, pp. 653--663.

\bibitem{soud2024fly}
M.~Soud, G.~Liebel, M.~Hamdaqa, A fly in the ointment: an empirical study on
  the characteristics of ethereum smart contract code weaknesses, Empirical
  Software Engineering 29~(1) (2024) 13.

\bibitem{zhang2023demystifying}
Z.~Zhang, B.~Zhang, W.~Xu, Z.~Lin, Demystifying exploitable bugs in smart
  contracts, ICSE, 2023.

\bibitem{parity-multisig-bug}
L.~Breidenbach, P.~Daian, A.~Juels, E.~Gün~Sirer, An in-depth look at the
  parity multisig bug,
  \url{http://hackingdistributed.com/2017/07/22/deep-dive-parity-bug/} (july
  2017).

\bibitem{parity-accidentally-kill}
{An Ethereum User}, Anyone can kill your contract,
  \url{https://github.com/openethereum/parity-ethereum/issues/6995}, November
  6, 2017 (november 2017).

\bibitem{thorchain}
A.~Hetman, {Unboxing tx.origin. Rune Token case},
  \url{https://www.adrianhetman.com/unboxing-tx-origin/}, Visited 2022-03-31
  (jul 2021).

\bibitem{powh-underflow}
E.~Banisadr, How \$800k evaporated from the powh coin ponzi scheme overnight,
  \url{https://blog.goodaudience.com/how-800k-evaporated-from-the-powh-coin-ponzi-scheme-overnight-1b025c33b530}
  (2018).

\bibitem{bec-batch-overflow}
p0n1, A disastrous vulnerability found in smart contracts of beautychain (bec),
  \url{https://medium.com/secbit-media/a-disastrous-vulnerability-found-in-smart-contracts-of-beautychain-bec-dbf24ddbc30e}
  (2018).

\bibitem{icon-fatal}
r/CryptoCurrency, All transfers of the \$800m marketcap icon (icx) are
  completely disabled due to fatal bug in smart contract,
  \url{https://www.reddit.com/r/CryptoCurrency/comments/8rfm9w/all_transfers_of_the_800m_marketcap_icon_icx_are/}
  (2016).

\bibitem{barboni2021sumo}
M.~Barboni, A.~Morichetta, A.~Polini, Sumo: A mutation testing strategy for
  solidity smart contracts (2021) 50--59.

\bibitem{chapman2019deviant}
P.~Chapman, D.~Xu, L.~Deng, Y.~Xiong, Deviant: A mutation testing tool for
  solidity smart contracts, in: International Conference on Blockchain
  (Blockchain), IEEE, 2019, pp. 319--324.

\bibitem{honig2019practical}
J.~J. Honig, M.~H. Everts, M.~Huisman, Practical mutation testing for smart
  contracts, in: Data Privacy Management, Cryptocurrencies and Blockchain
  Technology, Springer, 2019, pp. 289--303.

\bibitem{li2019musc}
Z.~Li, H.~Wu, J.~Xu, X.~Wang, L.~Zhang, Z.~Chen, Musc: A tool for mutation
  testing of ethereum smart contract, in: 2019 34th IEEE/ACM international
  conference on automated software engineering (ASE), IEEE, 2019, pp.
  1198--1201.

\bibitem{hartel2020mutation}
P.~Hartel, R.~Schumi, Mutation testing of smart contracts at scale, in:
  International Conference on Tests and Proofs, Springer, 2020, pp. 23--42.

\bibitem{andesta2020testing}
E.~Andesta, F.~Faghih, M.~Fooladgar, Testing smart contracts gets smarter, in:
  2020 10th International Conference on Computer and Knowledge Engineering
  (ICCKE), IEEE, 2020, pp. 405--412.

\bibitem{li2018targeted}
S.~Li, F.~Hariri, G.~Agha, Targeted test generation for actor systems, in: 32nd
  European Conference on Object-Oriented Programming (ECOOP),
  Schloss-Dagstuhl-Leibniz Zentrum f{\"u}r Informatik, 2018.

\bibitem{jensen2013automated}
C.~S. Jensen, M.~R. Prasad, A.~M{\o}ller, Automated testing with targeted event
  sequence generation, in: Proceedings of the International Symposium on
  Software Testing and Analysis, 2013, pp. 67--77.

\bibitem{arzt2015using}
S.~Arzt, S.~Rasthofer, R.~Hahn, E.~Bodden, Using targeted symbolic execution
  for reducing false-positives in dataflow analysis, in: Proceedings of the 4th
  ACM SIGPLAN International Workshop on State of the Art in Program Analysis,
  2015, pp. 1--6.

\bibitem{manticore-sample-codes}
ManticoreCommunity,
  \href{https://github.com/trailofbits/manticore/tree/master/examples/evm}{Manticore
  examples - evm} (2023).
\newline\urlprefix\url{https://github.com/trailofbits/manticore/tree/master/examples/evm}

\bibitem{slither-trophies}
Crytic, slither/trophies.md at master · crytic/slither,
  \url{https://github.com/crytic/slither/blob/master/trophies.md} (2023).

\bibitem{zhang2019mpro}
W.~Zhang, S.~Banescu, L.~Pasos, S.~Stewart, V.~Ganesh, Mpro: Combining static
  and symbolic analysis for scalable testing of smart contract, in: IEEE 30th
  International Symposium on Software Reliability Engineering (ISSRE), IEEE,
  2019, pp. 456--462.

\bibitem{yu2020smart}
X.~L. Yu, O.~Al-Bataineh, D.~Lo, A.~Roychoudhury, Smart contract repair, ACM
  Transactions on Software Engineering and Methodology (TOSEM) 29~(4) (2020)
  1--32.

\bibitem{wood2014ethereum}
G.~Wood, et~al., Ethereum: A secure decentralised generalised transaction
  ledger, Ethereum project yellow paper 151~(2014) (2014) 1--32.

\bibitem{ply}
D.~Beazley, \href{https://www.dabeaz.com/ply/ply.html}{Ply (python lex-yacc)}
  (2001).
\newline\urlprefix\url{https://www.dabeaz.com/ply/ply.html}

\bibitem{z3}
L.~De~Moura, N.~Bj{\o}rner, Z3: An efficient smt solver, in: International
  conference on Tools and Algorithms for the Construction and Analysis of
  Systems, Springer, 2008, pp. 337--340.

\bibitem{hardhat}
H.~Team, Hardhat | ethereum development environment for professionals,
  \url{https://hardhat.org/} (2024).

\bibitem{Driessen2021}
S.~W. Driessen, D.~D. Nucci, G.~Monsieur, D.~A. Tamburri, W.-J. {Van Den
  Heuvel}, {Automated Test-Case Generation for Solidity Smart Contracts: the
  AGSolT Approach and its Evaluation}, arXiv preprint arXiv:2102.08864v2
  (2021).

\bibitem{olsthoorn2022syntest}
M.~Olsthoorn, D.~Stallenberg, A.~van Deursen, A.~Panichella, Syntest-solidity:
  Automated test case generation and fuzzing for smart contracts, in: The 44th
  International Conference on Software Engineering-Demonstration Track,
  IEEE/ACM, 2022.

\end{thebibliography}

\end{document}